\documentclass[aps,pre,onecolumn,showpacs,superscriptaddress,groupedaddress,rmp]{revtex4-2}
\usepackage{scrextend}
\usepackage[toc,page]{appendix}
\usepackage{verbatim}
 \usepackage{graphicx}
 \usepackage{hyperref}
 \usepackage{mathtools}
 \usepackage{bm}
 \usepackage{lipsum}
 \allowdisplaybreaks
 \usepackage{amsmath}
 
 \usepackage{xcolor}
 \usepackage[makeroom]{cancel}
 \usepackage{subcaption}
 
 \usepackage{amsmath}

\begin{document}
	\title{ Accuracy in readout of  glutamate concentrations by neuronal cells.  }
 	\author{Swoyam Biswal   }
 	\email{swooropa19221104@iitgoa.ac.in }
 	\author{Vaibhav Wasnik}
 		\email{wasnik@iitgoa.ac.in}
 
 	 \affiliation{Indian Institute of Technology, Goa}

	\begin{abstract}

 	   Glutamate and glycine are  important neurotransmitters in the brain.  An action potential propagating in the terminal of a presynatic neuron causes the release of  glutamate and glycine    in the synapse by vesicles fusing with the cell membrane, which then  activate various receptors on the cell membrane of the  post synaptic neuron. Entry of $Ca^{2+}$ through the activated NMDA receptors leads to a host of cellular processes of which long term potentiation is of crucial importance because it is widely considered to be one of the major mechanisms behind learning and memory.  
 	 By analysing      the  readout of    glutamate   concentration by the post synaptic neurons during $Ca^{2+}$ signaling, we   find that  the average receptor density in hippocampal neurons has evolved to allow for accurate measurement of the  glutamate   concentration in the synaptic cleft.
		\end{abstract}

	\maketitle
	\section{Introduction}
	Cells  have evolved to be exceptional      information processing machines.    \emph{E Coli} can detect concentrations of 3.2 nM of the
	attractant aspartate  which is equivalent to around three molecules in the cell volume \cite{expt1},\cite{wingreenendres}. Receptors in the retina can detect a single photon \cite{retina}.  Eukaryotic cells are	  known to measure and respond to extremely shallow	gradients of chemical signals \cite{gradient1},\cite{gradient2},\cite{gradient3}. Understanding of limitations to cellular measurement  theoretically was first carried out in the  work of Berg and Purcell \cite{bergpurcell} who showed that  the chemotatic sensitivity of EColi approaches that  allowed by optimal design.   Since then theoretical works have studied various aspects of the problem, from the role of receptor kinetics and receptor cooperativity \cite{bialeksetyasger},\cite{wangetal} in concentration measurements, to  reduction in noise in   concentration measurements because of cellular communication   \cite{communications}, to  limitation in measurement of    temporal concentration changes  \cite{temporal}.  Even limitations to  the measurement of cellular gradients were    considered in a host of works \cite{gradients11}-\cite{gradients77}.

	The list of theoretical works mentioned above is far from exhaustive, however majority of  such  studies  in literature   have tried to understand the problem of limitations to cellular measurements by reducing the cell to a spherical object with  measurements done by cell surface receptors  without any   reference to the activities in the cellular cytoplasm. However the processing of extracellular signals is done through the reactions happening in the cellular cytoplasm. Understanding limitations to cellular measurements  carried out using   reactions happening in the cellular cytoplasm as readouts is relatively unexplored in literature.   In neuronal cells the problem of    limitations on measurements of neurotransmitter concentrations has not been studied theoretically despite its importance given that neurons communicate using the neurotransmitters that are   released in the synapse.  In the post synaptic neurons   the membrane   potential reaching a threshold value   causes   $Na^+$ channels on the membrane to open up resulting in a    substantial influx of $Na^+$  ions into the cells leading to the depolarizing phase of  an action potential that leads to  the membrane potential shooting up.
	 In the depolarizing phase   $Ca^{2+}$ also enters the post synaptic neuron   through activation of the NMDA  receptors. This $Ca^{2+}$ attaches to  calmodulin in the cytoplasm which then attaches to kinases  including CaMKII causing their activation. Activated CaMKII phosphorylates AMPA receptors thereby increasing the conduction of sodium ions. It also increases the movement of AMPA receptors to the neuronal membrane thereby increasing the amount of $Na^+$  that could move inside the neuronal cell.  This leads to synaptic enhancement and leads to long term potentiation, a important cellular mechanism that underlies learning and memory. In order for this process to be robust, it is penultimate that synaptic enhancement be linked to the strength of the action potential in the pre-synaptic neuron.  As the action potential subsides  the $Na^+$  and $K^+$ channels are  closed and   the corresponding ion concentrations in the neuron are set back to the resting stage.  The  NMDA receptor   gets activated when two  glutamate and two glycine ligands are attached to it.     Excessive glutamate release and over expression of NMDAR's has been linked to NMDAR-dependent neurotoxicity 
	 in several CNS disorders, including ischaemic stroke 
	 and neurodegenerative disorders such as Parkinson
	 disease, Alzheimer disease and Huntington disease, while  less than optimal glutamate release has been linked to depression and other psychiatric disorders\cite{neurotransmitter}.  It is hence natural that the neurons should be employed with mechanisms that detect the glutamate concentrations with appropriate accuracy.     In this work, we     evaluate the limitations to measurement of glutmate     concentration in neuronal cells and show that the average receptor density in hippocampal neurons is   poised  to allow for optimal measurement of these concentrations.

	\section*{Theory}

	
	\begin{figure}[h!]
\includegraphics[scale=.35]{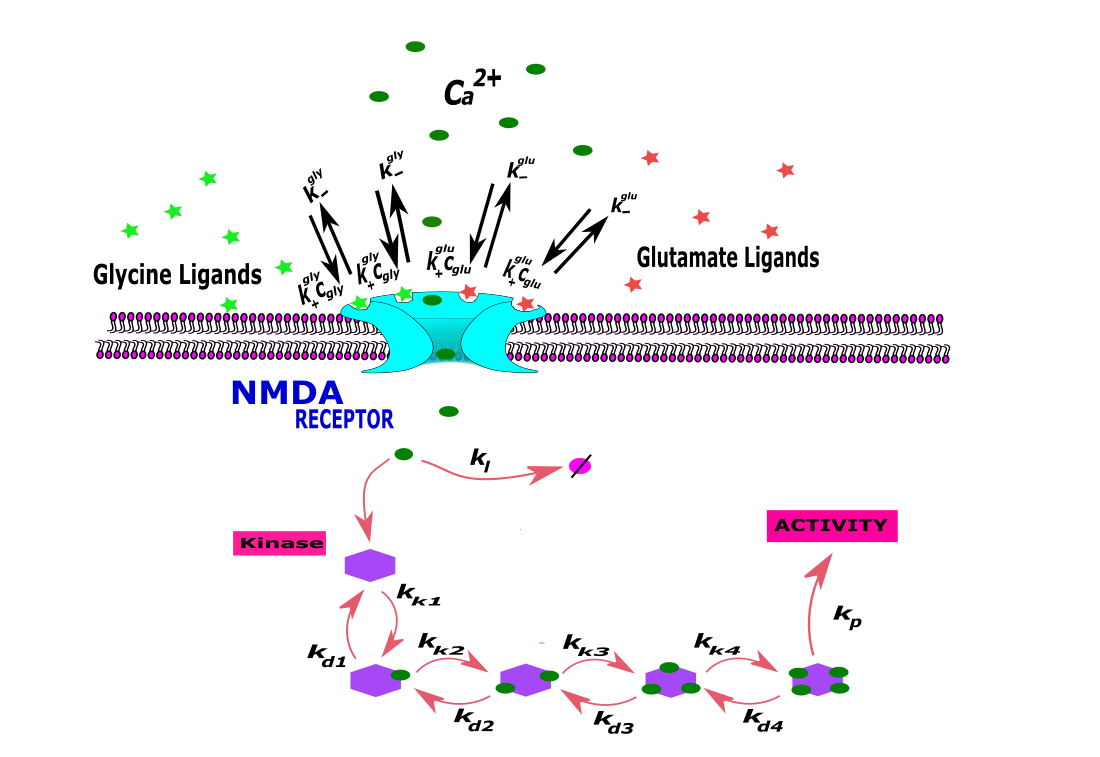}
\caption{  
	Schematic representation of elements of  the $Ca^{2+}$  signal transduction pathway that are studied in the paper. The rate equations for the reactions  are presented in Eq.\ref{basic equations}.  }
\label{schemata}
	\end{figure}

 In order to consider the readout of the     glutamate concentration by the neuron  consider the schemata of reactions in Fig.\ref{schemata}. The concentration of the glutamate/glycine ligands  is represented by $c_{glu}/c_{gly}$. Two  glutamine and glycine ligands are required to attach to the NMDA receptor in order for it to open to $Ca^{2+}$ influx. The rate of attachment of the ligand to the NMDA receptor is denoted as  $k_+^{glu}c_{glu} /k_+^{gly}c_{gly}$ and the rate of detachment is $k_-^{glu}/k_-^{gly}$.  The attachment of the ligands to the  NMDA receptors opens up   $Ca^{2+}$ channel in the receptor leading to influx of $Ca^{2+}$, whose concentration inside the cell is represented by $C_{Ca}(t)$. The $Ca^{2+}$ then attaches to Calmodulin  which has four calcium binding sites organized into two globular domains.  The C terminal lobe contains     two high-affinity $Ca^{2+}$ -binding sites, while the N-terminal lobe contains two sites with lower $Ca^{2+}$  affinity. Since we expect the attachment of calcium to these domains to happen sequentially, we model the attachment of the first $Ca^{2+}$ to calmoudlin  to produce a $K-Ca$ at the rate $k_{k1}$. The detachment rate of $Ca^{2+}$ from $K-Ca$ is taken to be $k_{d1}$.  We model the attachment of the   $Ca^{2+}$ to  $K-Ca$  to produce a $K-Ca-Ca$ at the rate $k_{k2}$. The detachment rate of $Ca^{2+}$ from $K-Ca-Ca$ is taken to be $k_{d2}$. We similarly define $k_{k3},k_{k4}$ and $k_{d3}, k_{d4}$.  The   $K-Ca-Ca-Ca-Ca$ complex phosphorylates AMPA receptors as well as increases the movement of AMPA receptors to the plasma membrane. We model this activity    happening at a rate $k_{p}$, the amount  of which we label by $C_{Pr}$. Let $k_{dp}$ be the rate at which this activity gets reduced. Finally the $Ca^{2+}$ coming into the cell has to also be removed from the cell. This happens at a rate $k_l$.  We denote by $N$ the rate at which $Ca^{2+}$, enters the cell when the $Ca^{2+}$ channel is open.   Assume that the ligands attach the  receptor   at  times $t^j_{2i-1}$ and detach at times $t^j_{2i}$, with $T$ representing  the time duration of the action potential which is the measurement time and $ j = 1,2,3,4$ correspond to the first glutamine ligand, the second glutamine ligand, the first glycine ligand and the second glycine ligand respectively. Also let us assume the probability of realizing these attachment detachment events is presented by $P(\{t_i^j\})$.  We have the following rate equations
 
 \begin{equation}\label{basic equations}
 \begin{split}
 \frac{dC_{Ca}}{dt}  &=   N \Pi_{j=1,4}[   \sum_{  t^j_{i}< T } (-1)^{i-1}  \Theta(t-t^j_{i}) ]-(k_l+k_{k1} C_K + k_{k2} C_{K-Ca} + k_{k3} C_{K-Ca-Ca} + k_{k4} C_{K-Ca-Ca-Ca} ) C_{Ca} +\\ & k_{d1}C_{K-Ca} 
+k_{d2}C_{K-Ca-Ca} +k_{d3}C_{K-Ca-Ca-Ca}   +k_{d4}C_{K-Ca-Ca-Ca-Ca}  \\
 \frac{dC_{K}}{dt}&=  k_{d1} C_{K-Ca}  - k_{k1} C_{Ca} C_K   \\
 \frac{dC_{K-Ca}}{dt}&= k_{k1} C_{Ca} C_K +k_{d2} C_{K-Ca-Ca}  - k_{d1} C_{K-Ca}  \\
 \frac{dC_{K-Ca-Ca}}{dt}&= k_{k2} C_{K-Ca} C_{Ca} + k_{d3} C_{K-Ca-Ca-Ca} - k_{d2} C_{K-Ca-Ca}  \\
 \frac{dC_{K-Ca-Ca-Ca}}{dt}&= k_{k3} C_{K-Ca-Ca} C_{Ca} +  k_{d4} C_{K-Ca-Ca-Ca-Ca} - k_{d3} C_{K-Ca-Ca-Ca}  \\
 \frac{dC_{K-Ca-Ca-Ca-Ca}}{dt}&= k_{k4} C_{K-Ca-Ca-Ca} C_{Ca} - k_{d4} C_{K-Ca-Ca-Ca-Ca}  \\
 \frac{dC_{Pr}}{dt}&= k_{p} C_{K-Ca-Ca-Ca-Ca} - k_{dp} C_{Pr}\\ 
 \end{split}
 \end{equation}
 where, $\Theta(x)$ is   
 \begin{equation*}
 \Theta(x) = \begin{cases}
 1, \;     x > 0, \\
 0, \;   x  < 0.
 \end{cases}
 \end{equation*}

The association and dissociation rate constant of glutamate with the NMDA receptor are \cite{NMDA} $k+ \approx 10^7 M^{-1} s^{-1}$, $k_- \sim 8 s^{-1}$ respectively. The concentration of glutamate in the synaptic cleft is also understood to be of the order of a few millimolars \cite{glutamate concentration}. As such one would have expected the probability of occupancy of a NMDA receptor to be $\frac{k_+ c}{k_+ c + k_-} \approx 1$. However, the glutamate concentration   rapidly diffuses after arriving at the synaptic cleft through a vesicle. The  decay is  exponential  with a time constant $\sim 3 ms$.   \cite{Holmes} found that because of this rapid diffusion, for the range of glutamate molecules in a vesicle $(1500-4000)$ \cite{riverois}, \cite{tim}  and value of most likely diffusion constant $D = .25 \mu m^2/s$ the receptor occupancy (with two glutamate molecules) was $30 -81 $ percent. Glycine concentration in the synaptic cleft would also be expected to similarly diffuse out on similar time scales and given that the rate constants are of a similar magnitude \cite{glycine rates} would lead to similar receptor occupancy.  The probability of receptor occupancy (we are considering either glutamate or glycine here) obeys the equation

\begin{eqnarray}
\frac{dp_2}{dt}&&= k_+ c p_1-k_- p_2\nonumber\\
\frac{dp_1}{dt}&&= k_+ c(1-p_1-p_2)-k_- p_1 = k_+ c -(k_+ c + k_-)p_1 - k_+c p_2\nonumber\\
\end{eqnarray}

where $p_1/p_2$ is the probability of the receptor being occupied with one/two glutamate (or glycine) molecule(s). If we define $\frac{k_+c}{k_-} = k$ and $k_- t = t'$. The above can be written as 
\begin{eqnarray}
\frac{dp_2}{dt'}&&= k p_1- p_2\nonumber\\
\frac{dp_1}{dt'}&&\approx   k -(k+1) p_1 - k  p_2 \nonumber
\end{eqnarray}
or
\begin{eqnarray}
\frac{1}{k}(\frac{d^2p_2}{dt'^2} + \frac{dp_2}{dt'}) &&\approx   k - \frac{k+1}{k}\frac{dp_2}{dt'} - \frac{k+1}{k} p_2 - k  p_2 \nonumber
\end{eqnarray}
which gives assuming $k>>1$, gives
\begin{eqnarray}
  k^2 -(k+1) \frac{dp_2}{dt'} - (k^2+k+1)  p_2 \approx 0
  \label{approx}
\end{eqnarray}
implying
\begin{eqnarray}
p_2(T)= \frac{k^2 }{k^2+k+1} \left(1-e^{-\frac{\left(k^2+k+1\right) (k_-T)}{k+1}}\right)
\end{eqnarray}
We see that plugging $T = 3 ms$ gives $p(T)\sim 1$. Diffusion removing the glutamate out of the synaptic cleft, however implies that we have to consider a renormalized value of $k_+c$ so as to get $p(T)$ between $.30-.81$, hence the renormalized values of $k_+c$ should be such that  $k_+ cT \in [ .4, 1.7]$. Note that $\frac{k_+c}{k_-}$ is still $>>1$ in this range and since   this approximation was used above,     Eq.\ref{approx} is still valid. 
Now going back to Eq.\ref{basic equations}, if we were to assume  that the measurement time to be so short   that,
 \begin{enumerate} 
  \item It only  resulted in only $2$ glutamine and $2$ glycine attachment events, such that $t_7$ is the time of the last attachment event, with no detachment event.
  \item  Except for the rate equation for $C_{K}$ the first term on the R.H.S is the  most dominant   in every equation,
\end{enumerate}
    we would have
\begin{equation}\label{assumption}
\begin{split}
C_{Ca}(t)& = N(t-t_7)\\
C_{K-Ca}(t)&= N k_{k1}  C_K \frac{ (t-t_7)^2}{2} \\
C_{K-Ca-Ca}(t)&= N^2 k_{k1} k_{k2}  C_K \frac{ (t-t_7)^4}{2\times 4}   \\
C_{K-Ca-Ca-Ca}(t)&= N^3 k_{k1} k_{k2} k_{k3}  C_K \frac{ (t-t_7)^6}{2\times 4\times 6}   \\
C_{K-Ca-Ca-Ca-Ca}(t)&= N^4 k_{k1} k_{k2} k_{k3} k_{k4}  C_K \frac{ (t-t_7)^8}{2\times 4\times 6\times 8}   \\
C_{Pr} &= N^5 k_{k1} k_{k2} k_{k3} k_{k4} k_{p} C_{K }  \frac{ (t-t_7)^{9}}{2\times 4 \times 6 \times 8 \times 9}   
\end{split}
\end{equation}
Now, at rest  most neurons have an intracellular calcium concentration of about $50 - 100$ nM that can rise transiently during electrical activity to levels that are 10 to 100 times higher \cite{lipp} in around a millisecond. We can hence assume that $N = 10000 nM/  ms =10^{-2}   M s^{-1}$. From \cite{calmodulin rate constants} we find   $k_{d1}, k_{d2} = 500$ $s^{-1}$, $k_{d3}, k_{d4} = 6$ $s^{-1}$.    An estimate of  calmodulin concentration used in literature is $C_K = 10^{-4}$M \cite{calmodulin concentration}. The association rate of $Ca^{2+}$  with calmodulin taken from literature \cite{calmoudlin association rate} is between $k_{k3}, k_{k4} =  6.8\times  10^{6}\;   M^{-1} s^{-1}$ and $k_{k1}, k_{k2} = 108 \times  10^{6}\;   M^{-1} s^{-1}$. Hence
   \begin{equation} \label{proof}
   \begin{split}
   C_K(t)&\sim 10^{-4} M\\
   C_{Ca}(t)& \sim  10^{-2} \times 10^{-3} = 10^{-5}M\\
   C_{K-Ca}(t)&\sim 10^{-2}\times  (100\times 10^6)  \times  10^{-4} \times 10^{-6} M = 10^{-4}   M\\
   C_{K-Ca-Ca}(t)&\sim 10^{-4} \times  (100\times 10^6)^2   \times 10^{-4} \times    10^{-12} \times 10^{-1}M =    10^{-5}   M \\
    C_{K-Ca-Ca-Ca}(t)&\sim 10^{-6}\times   (10\times 10^6) \times(100\times 10^6)^2   \times 10^{-4} \times    10^{-18} \times 10^{-1} M= 10^{-6} M\\
     C_{K-Ca-Ca-Ca-Ca}(t)&\sim 10^{-8}  \times (10\times 10^6)^2 \times(100\times 10^6)^2  \times   10^{-4} \times    10^{-24} \times 10^{-2} M=  10^{-8} M\\
   \end{split}
   \end{equation}

\begin{figure}[h!]
 	\includegraphics[scale=.65]{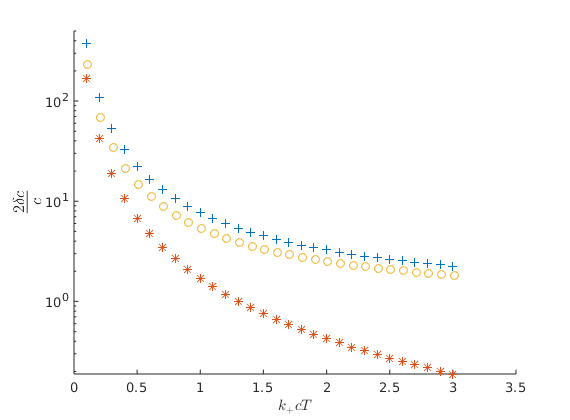}
 	\caption{ (a )Error in evaluation of glutamate and glycine concentration together by a single receptor, evaluated using Eq.\ref{evaluation_n} represented by  $'*'$ and the error evaluated  using Eq.\ref{evaluation} represented by $'\circ'$ are indeed close as mentioned in the text.     Error evaluated using  Eq.\ref{approximation} is represented by $'+'$. One can see close agreement with exact error for smaller values of $k_+ c T$ as expected.(b) Error in measurement of glutamate concentration by the cell as a function of glutamate molecules in a synaptic vesicle. One can see that when this number becomes around $1500$, that the error starts becoming $.5$  }
 	\label{fig1}
 \end{figure}

  The $\sim$ above implies an order of magnitude estimate.  Note that any of the $k_{k1}, k_{k2}, k_{k3}, k_{k4}$ multiplied by $C_{Ca}(t)$ gives a factor of atleast $10^2$. Also from above we see that addition of  every $Ca^{2+}$ to a molecule reduces its concentration atleast by a factor of $10$. This then justifies the claim that  the first term on the R.H.S is the most dominant on the R.H.S in every equation and hence our assumption in derivation of  Eq.\ref{assumption} is consistent. This is also  seen in Fig.\ref{fig1} where error evaluated assuming $C_{Cpr}$ from Eq.\ref{assumption} is similar to error evaluated using $C_{Cpr}$ obtained from Eq.\ref{basic equations}. 
   
   We note that if we were to   associate the rates $108 \times  10^{6}\;   M^{-1} s^{-1} $, $ 6.8\times  10^{6}\;   M^{-1} s^{-1}$ to a different permutation of  $k_{k1}, k_{k2} , k_{k3}, k_{k4}$ (for e.g.  $k_{k2}, k_{k4} =  6.8\times  10^{6}\;   M^{-1} s^{-1}$ and $k_{k1}, k_{k3} = 108 \times  10^{6}\;   M^{-1} s^{-1}$)with the corresponding permutations of  $k_{d1}, k_{d2} , k_{d3}, k_{d4}$, we would still have the first term  on the R.H.S of rate equations of Eq.1, except for the rate equation for $C_{K}$, to be dominant. This would then cause the error in measurement of glutamate concentration to be independent of the reaction rates in the cytoplasm. Hence as far as the problem of evaluation of the  value of  error in concentration measurement goes, the order of attachment of glutamate to the NMDA receptor is irrelevant.

To see this, let us for example consider the equation 
\begin{eqnarray*}
\frac{dC_{K-Ca-Ca-Ca}}{dt}&= k_{k3} C_{K-Ca-Ca} C_{Ca} +  k_{d4} C_{K-Ca-Ca-Ca-Ca} - k_{d3} C_{K-Ca-Ca-Ca}  \\
\end{eqnarray*}
If we consider ratio of terms on the right hand side, we get
\begin{eqnarray*}
k_{k3} C_{K-Ca-Ca} C_{Ca}: k_{d4} C_{K-Ca-Ca-Ca-Ca} : k_{d3} C_{K-Ca-Ca-Ca} &&=\nonumber\\
k_{k3}  C_{Ca}  :  k_{d4} \frac{C_{K-Ca-Ca-Ca-Ca}}{ C_{K-Ca-Ca}} :  k_{d3} \frac{C_{K-Ca-Ca-Ca}}{ C_{K-Ca-Ca}} &&=\nonumber\\
k_{k3}C_{Ca} :  \frac{ k_{d4}k_{k3} k_{k4} N^2  (t-t_7)^4 }{6\times 8} : \frac{k_{d3} k_{k3}    N(t-t_7)^2  }{6}&&=\nonumber\\
k_{k3}C_{Ca} :  \frac{ k_{d4}k_{k3}C_{Ca}  k_{k4} N (t-t_7)^3 }{6\times 8} : \frac{k_{d3} k_{k3}C_{Ca}     (t-t_7) }{6}&&=\nonumber\\
1 :  \frac{ k_{d4} k_{k4} N   (t-t_7)^3 }{6\times 8} : \frac{k_{d3}      (t-t_7)   }{6}&&=\nonumber\\ 
\end{eqnarray*}

where the second ,third,fourth lines use Eq.5 in the text. Given that $N    =10^{-2}   M s^{-1}$ and $(t-t_7)$ is of the order of milliseconds, we can easily see that no matter what values we choose for the rate constants, the first term in rate equation for $C_{K-Ca-Ca-Ca} $ is the dominant one. Such arguments can be made for every rate equation in Eq.1, except the equation for $C_K$.

The calcium channel will only open when $2$ glutamate and $2$ glycine molecules are attached to the receptor. Let us assume that these attachment events happen at times $t_1 < t_3< t_5< t_7 < T$. Calcium influx happens after time $t_7$. The probability of these remaining attached till time $T$ is 
\begin{eqnarray}
P(t_1,t_3,t_5,t_7)&&= [ k_+^{gly}c_{gly}]^2dt_1 dt_3 [ k_+^{glu}c_{glu}]^2dt_5 dt_7 e^{-k_+^{gly}c_{gly} t_1}e^{-k_-^{gly}(T-t_1)} e^{-k_+^{gly}c_{gly} t_3}e^{-k_-^{gly}(T-t_3)}e^{-k_+^{glu}c_{glu} t_5}e^{-k_-^{glu}(T-t_5)}\nonumber\\ &&e^{-k_+^{glu}c_{glu} t_7}e^{-k_-^{glu}(T-t_7)}\nonumber\\
\end{eqnarray}

 If any of the ions detaches the calcium influx stops. The calcium influx only starts after the respective ion reattaches. Since the  time interval of measurement $T$ is so small that $k_-^{gly}T <<1$ and $k_-^{glu}T <<1$, the probability of this and subsequents detachment and reattachment happening  are miniscule compared to $P(t_1,t_3,t_5,t_7)$. Hence,
 \begin{eqnarray}
 \langle C_{Pr} \rangle&& = \sum_{Permutate:\tilde{t}_1,\tilde{t}_3,\tilde{t}_5,\tilde{t}_7} \int_{\tilde{t}_1< \tilde{t}_3<\tilde{t}_5<\tilde{t}_7<T} [ k_+^{gly}c_{gly}  k_+^{glu}c_{glu}]^2d\tilde{t}_1 d\tilde{t}_3  d\tilde{t}_5 d\tilde{t}_7 e^{-k_+^{gly}c_{gly} \tilde{t}_1} e^{-k_+^{gly}c_{gly} \tilde{t}_3} e^{-k_+^{glu}c_{glu} \tilde{t}_5}  e^{-k_+^{glu}c_{glu} \tilde{t}_7}\nonumber\\ &&C_{Pr}(T,g(\tilde{t}_1,\tilde{t}_3,\tilde{t}_5,\tilde{t}_7))\nonumber\\
 \langle C^2_{Pr} \rangle&& = \sum_{Permutate:\tilde{t}_1,\tilde{t}_3,\tilde{t}_5,\tilde{t}_7}\int_{\tilde{t}_1< \tilde{t}_3<\tilde{t}_5<\tilde{t}_7<T} [ k_+^{gly}c_{gly}  k_+^{glu}c_{glu}]^2d\tilde{t}_1 d\tilde{t}_3  d\tilde{t}_5 d\tilde{t}_7 e^{-k_+^{gly}c_{gly} \tilde{t}_1} e^{-k_+^{gly}c_{gly} \tilde{t}_3} e^{-k_+^{glu}c_{glu} \tilde{t}_5}  e^{-k_+^{glu}c_{glu} \tilde{t}_7}  \nonumber\\
 &&C^2_{Pr}(T,g(\tilde{t}_1,\tilde{t}_3,\tilde{t}_5,\tilde{t}_7))\nonumber\\
 \label{eqns1}
 \end{eqnarray}
where here $g(\tilde{t}_1,\tilde{t}_3,\tilde{t}_5,\tilde{t}_7)$ is the greatest among $ \tilde{t}_1,\tilde{t}_3,\tilde{t}_5,\tilde{t}_7$. For calculational simplicity let us say that $k_+^{glyc} c_{gly} = k_+^{glu} c_{glu} = k c$. Now,  
 \begin{eqnarray}
 \frac{\delta c}{ c} = \frac{\sqrt{\langle C_{Pr}^2 \rangle -\langle C_{Pr} \rangle^2  }}{c\frac{d \langle C_{Pr} \rangle}{dc}}
 \label{evaluation}
 \end{eqnarray}
   We  hence have with $n =9 $ and $ C= \frac{   N^5 k_{k1} k_{k2} k_{k3} k_{k4} k_{p} C_{K } } {2\times 4 \times 6 \times 8 \times 9}    $ below
 \begin{eqnarray}
 \langle C_{Pr} \rangle_n=&& \sum_{Permutate:\tilde{t}_1,\tilde{t}_3,\tilde{t}_5,\tilde{t}_7} \int_{\tilde{t}_1< \tilde{t}_3<\tilde{t}_5<\tilde{t}_7<T} [ k_+^{gly}c_{gly}  k_+^{glu}c_{glu}]^2d\tilde{t}_1 d\tilde{t}_3  d\tilde{t}_5 d\tilde{t}_7 e^{-k_+^{gly}c_{gly} \tilde{t}_1} e^{-k_+^{gly}c_{gly} \tilde{t}_3} e^{-k_+^{glu}c_{glu} \tilde{t}_5}  e^{-k_+^{glu}c_{glu} \tilde{t}_7}\nonumber\\&& C(T-\tilde{t}_7)^n\nonumber\\
 \langle C^2_{Pr} \rangle_n =&&\sum_{Permutate:\tilde{t}_1,\tilde{t}_3,\tilde{t}_5,\tilde{t}_7}\int_{\tilde{t}_1< \tilde{t}_3<\tilde{t}_5<\tilde{t}_7<T} [ k_+^{gly}c_{gly}  k_+^{glu}c_{glu}]^2d\tilde{t}_1 d\tilde{t}_3  d\tilde{t}_5 d\tilde{t}_7 e^{-k_+^{gly}c_{gly} \tilde{t}_1} e^{-k_+^{gly}c_{gly} \tilde{t}_3} e^{-k_+^{glu}c_{glu} \tilde{t}_5}  e^{-k_+^{glu}c_{glu} \tilde{t}_7}\nonumber\\&& C^2(T-\tilde{t}_7)^{2n}\nonumber\\
\end{eqnarray}
or,
\begin{eqnarray}
\langle C_{Pr}\rangle_n= &&\frac{1}{(kc)^n} \sum_{Permutate:\tilde{t}_1,\tilde{t}_3,\tilde{t}_5,\tilde{t}_7} \int_{\tilde{t}_1< \tilde{t}_3<\tilde{t}_5<\tilde{t}_7< kcT} d\tilde{t}_1 d\tilde{t}_3  d\tilde{t}_5 d\tilde{t}_7 e^{- \tilde{t}_1} e^{- \tilde{t}_3} e^{-  \tilde{t}_5}  e^{-  \tilde{t}_7} C(kc T-\tilde{t}_7)^n\nonumber\\
&&=\frac{4! C}{6 (kc)^n} \int_0^{kcT}  e^{ -\tilde{t}_7 }\left(1-e^{ -\tilde{t}_7} \right)^3(kc T-\tilde{t}_7)^n d\tilde{t}_7 \nonumber\\
 \langle C_{Pr}^2 \rangle_n&& =  \frac{4! C^2}{6 (kc)^{2n}} \int_0^{kcT}  e^{ -\tilde{t}_7 }   \left(1-e^{ -\tilde{t}_7 } \right)^3(kc T-\tilde{t}_7)^{2n} d\tilde{t}_7 
 \label{Tminust}
\end{eqnarray}
 
  \begin{eqnarray}
 (\frac{\delta c}{ c})_n = \frac{\sqrt{\langle C_{Pr}^2 \rangle_n -\langle C_{Pr} \rangle^2_n  }}{c\frac{d \langle C_{Pr} \rangle_n}{dc}}
 \label{evaluation_n}
 \end{eqnarray}
 
We see that the  error evaluated using  Eq.\ref{evaluation} is similar to error evaluated using Eq.\ref{evaluation_n}   as shown in Fig.\ref{fig1},  implying the assumptions made in getting Eq.\ref{assumption} were justified.

  For the range $kcT \in [.4,1.7]$,   the error in measurement of glutamate as well as glycine concentration   totals to $2\frac{\delta c}{c}\in[ 21.1-3.0]$. Measurements by multiple NMDA receptors would have to be done in order to ensure accuracy in concentration measurements.  If $N_{receptors}$ is the number of NMDA receptors per neuron, it would imply a reduction in error by a factor of $\sqrt{N_{receptors}}$. It is seen that \cite{receptor density} following synaptogenesis, functional AMPA and NMDA receptors are clustered  in the cultured hippocampal neurons with about  $\sim 400$ receptors/synapse, corresponding to $ 2\frac{\delta c}{c } \in [1.1, .15] $. The error in measurement of glutamate concentration on an average  should be  $ \frac{\delta c}{c } \in [0.5, .075] $ implying that the neurons can  atleast  detect concentration upto an error of $50$ percent. Since we would expect long term potentiation to be dependent upon whether the pre synaptic action potential is weak or strong, one would expect the neuron to atleast detect the glutamate concentration  with an accuracy which would decide the fate of the ease of  post synaptic neuron activation in the future.    
  Since error goes as $\frac{1}{\sqrt{N_{receptors}}}$, the error in concentration measurement should go as $\frac{\delta c}{c} \in [.5,.075]\times \sqrt{\frac{400}{N_{receptors}}}$. Hence if the receptor number was less than a factor of $10$ than what is seen phenomenologically (i.e 40), we would have $\frac{\delta c}{c} \in [1.5,.23]$, which implies that the error in measurement of concentration could be of the order of the ligand concentration and  hence the receptor number would  not be sufficient  to decipher if the incoming action potential of the pre synaptic neuron  was strong or weak, with the post synaptic neuron having a sizeable probability of reading a weak action potential of the pre synaptic neuron    as strong and vice versa. If we were to increase the receptor number by a factor of $10$ (i.e 4000) it would imply the error in concentration would go as $\frac{\delta c}{c} \in [.15,.023]$. Now we enter a regime, where at the least a $10$ percent error in concentration detection is obtained. Such a low level of error even though is acceptable does  not seem to be  a requirement in long term potentiation, where knowledge of a incoming action potential being strong or weak would be expected to be of  importance and  not   the actual value of the action potential.    It hence appears    that the number of NMDA receptors on the neuronal surface  have evolved as per the needs for long term potentiation.

  We could hence hypothesise, that the receptor number per neuron has been chosen by evolution to be apt for glutamate concentration detection. 
 
 One can proceed to evaluate an analytical form for the error by considering limit  $kcT << 1$. We see that
 
 \begin{eqnarray}
 \langle C_{Pr} \rangle_n&& = \frac{1}{(kc)^n} \sum_{Permutate:\tilde{t}_1,\tilde{t}_3,\tilde{t}_5,\tilde{t}_7} \int_{\tilde{t}_1< \tilde{t}_3<\tilde{t}_5<\tilde{t}_7< kcT} d\tilde{t}_1 d\tilde{t}_3  d\tilde{t}_5 d\tilde{t}_7 e^{- \tilde{t}_1} e^{- \tilde{t}_3} e^{-  \tilde{t}_5}  e^{-  \tilde{t}_7} C(kc T-\tilde{t}_7)^n\nonumber\\
 &&= \frac{4! C}{6 (kc)^n} \int_0^{kcT}  e^{ -\tilde{t}_7 }   \left(1-e^{ -\tilde{t}_7 } \right)^3(kc T-\tilde{t}_7)^n d\tilde{t}_7 \nonumber\\
 &&= \frac{4!C (kcT)^{n+1}}{6(kc)^n} \int_0^{1} e^{ -kcT x}    \left(1-e^{ -kcT x } \right)^3(1-x)^n dx \nonumber\\
 &&= \frac{4!C (kcT)^{n+1}}{6(kc)^n} \int_0^{1} e^{ -kcT  }  e^{ kcT y }   \left(1-e^{ -kcT  }e^{  kcT y } \right)^3y^n dy \nonumber\\
 \end{eqnarray}
  
 Similarly
 \begin{eqnarray}
 \langle C^2_{Pr} \rangle
   &&= \frac{4!C^2 (kcT)^{2n+1}}{6(kc)^{2n}} \int_0^{1} e^{ -kcT  }  e^{ kcT y }   \left(1-e^{ -kcT  }e^{  kcT y } \right)^3y^{2n} dy \nonumber\\
 \end{eqnarray}
 
As shown in the appendix for $kcT << 1$ we have 
 
 \begin{eqnarray}
 \frac{ \langle C^2_{Pr} \rangle_n-\langle C_{Pr} \rangle_n^2}{\langle C_{Pr} \rangle_n^2}&& =   	\frac{1}{4} \sqrt{  \frac{(n+1)(n+2)(n+3)^2(n+4)^2}{4!4(kcT)^{4}(2n+1)(2n+3)}-1}
 \label{approximation}
 \end{eqnarray}
 This is also plotted in Fig.\ref{fig1}. One can note  that this agrees well with the actual error for smaller values of $kcT$. 
 	
 \section*{Discussion}
 
   \cite{positional1}, \cite{positional2} have considered the problem of limitations to positional measurements in calcium signal transduction. However they considered the kinase concentration to be so high and non changing that the rate equations were considered to be linear. They also didn't consider the aspect that leads to non-linearities in the problem which arises because     calmodulin   gets activated only when four   $Ca^{2+}$ ions are attached to it. This simplified the form for errors in positional measurement obtained analytically under certain   assumptions. We have instead considered non-linearities in these equations and still could make some progress in analytical evaluations as in Eq.\ref{approximation}.We saw that Eq.\ref{assumption} were accurate enough in evaluation of error using Eq.\ref{evaluation}. The reason behind this was that the cytoplasmic rate constants as well as the measurement times colluded to produce order of magnitude concentration values of cytoplasmic reactants as illustrated in Eq.\ref{proof}, leading to   the most dominant terms in the rate equations Eq.\ref{basic equations}, being the first terms on the right hand side.   This also led to the error in concentration measurement being independent of the the cytoplasmic rate constants, despite the concentration of cytoplasmic reactants being dependent on them. We should note that it is the order of magnitude of these rate constants and the measurement time and not their absolute values that led to this phenomenon, implying that there was no specific fine tuning of these parameters needed by nature to lead to consistent accuracy in concentration measurement.    Since going by \cite{Holmes} we could assume that for both glycine and glutamate the renormalized values of $k_+^{glyc} c_{gly} T $ and $ k_+^{glu} c_{glu}T $ lie $\in [ .4, 1.7]$. Now, increasing $ k_+^{glyc} c_{gly} $ and/or  $k_+^{glu} c_{glu}$ would decrease the error, and since in our calculations    we   assumed that  $k_+^{glyc} c_{gly} = k_+^{glu} c_{glu} = k c$ and got that the glutamine is accurately detected for $kc T = .4$,  we can conclude that for  $k_+^{glyc} c_{gly} \neq k_+^{glu} c_{glu}$  with both  $k_+^{glyc} c_{gly} T$ and $ k_+^{glu} c_{glu}T$ lying $ \in [ .4, 1.7]$, the neuron would also accurately detect glutamine concentrations. As can be seen from Fig.\ref{fig1} around $1500$ glutamate molecules per vesicle (calculated using how $k_+^{glu} c_{glu} T$ varies with percentage receptor occupancy from Figure.$6$ in \cite{Holmes}) the error in glutamate concentration measurement by the cell becomes around $50$ percent. In regards to error reduction capabilities, we should be concerned with lowest concentration of glutamate measurable by the cell. As stated in \cite{riverois}, in rat brain cortex, number of glutamate molecules per synaptic vesicle was around $1100$ molecules, however  $30$ of
   all the SVs present operate with glutamate as a transmitter, this would raise the number of glutamate molecules per synaptic vesicle to $3640$. We can hence consider that a cell measuring  glutamate concentrations upto $50$ percent accuracy for     vesicles containing $1500$ molecules  to be very apt in measuring glutamate concentrations.

      It is always a question as to how various aspects of cellular constructions happened to evolve the way they did. In case of neuronal cells, one could question as to why cells have evolved to  be equipped with  the specific number of  cell surface receptors. As we had mentioned in the introduction, excessive as well as under optimum glutamate concentration in the synaptic cleft  would lead to several pathologies\cite{neurotransmitter}.  It is hence natural that the neurons should be employed with mechanisms that detect the glutamate concentrations accurately.  From our calculations it appears that the  way this is accomplished is by  using the specifically chosen number of NMDAR receptors on the neuronal surface that accurately detect concentration of glutamate.  We   have provided evidence as to why nature have chosen   the specific number of NMDAR receptors  on the surface of the neurons.

   \section*{Funding}
   We acknowledge funding from SERB Grant No:  EEQ/2021/000006.

\appendix
\section{ }
 We derive how Eq.\ref{approximation} is got in this appendix. We have 
  \begin{eqnarray}
  \langle C_{Pr} \rangle_n&& = \frac{1}{(kc)^n} \sum_{Permutate:\tilde{t}_1,\tilde{t}_3,\tilde{t}_5,\tilde{t}_7} \int_{\tilde{t}_1< \tilde{t}_3<\tilde{t}_5<\tilde{t}_7< kcT} d\tilde{t}_1 d\tilde{t}_3  d\tilde{t}_5 d\tilde{t}_7 e^{- \tilde{t}_1} e^{- \tilde{t}_3} e^{-  \tilde{t}_5}  e^{-  \tilde{t}_7} C(kc T-\tilde{t}_7)^n\nonumber\\
  &&= \frac{4! C}{6 (kc)^n} \int_0^{kcT}  e^{ -\tilde{t}_7 }   \left(1-e^{ -\tilde{t}_7 } \right)^3(kc T-\tilde{t}_7)^n dt_7
   \nonumber\\
  &&= \frac{4!C (kcT)^{n+1}}{6(kc)^n} \int_0^{1} e^{ -kcT x}    \left(1-e^{ -kcT x } \right)^3(1-x)^n dx \nonumber\\
  &&= \frac{4!C (kcT)^{n+1}}{6(kc)^n} \int_0^{1} e^{ -kcT  }  e^{ kcT y }   \left(1-e^{ -kcT  }e^{  kcT y } \right)^3y^n dy \nonumber\\
  &&= \frac{4!C (kcT)^{n+1}}{6(kc)^n} \int_0^{1}   e^{ kcT (y-1) }   \left(1-e^{  kcT (y-1) } \right)^3y^n dy \nonumber\\
  &&= \frac{4!C (kcT)^{n+1}}{6(kc)^n} \int_0^{1}   e^{ kcT (y-1) }   \left(1-e^{  3kcT (y-1) }-3e^{  kcT (y-1)}+3e^{ 2 kcT (y-1)} \right)y^n dy \nonumber\\
  &&= \frac{4!C (kcT)^{n+1}}{6(kc)^n} \int_0^{1}   \left( e^{ kcT (y-1) }  -e^{  4kcT (y-1) }-3e^{  2kcT (y-1)}+3e^{ 3 kcT (y-1)} \right)y^n dy \nonumber\\ 
  \end{eqnarray}
  Now  upto $\mathcal{O}((kcT)^3)$
  \begin{eqnarray}
  e^{kcT(y-1)}=1+kcT(y-1)+\frac{(kcT(y-1))^2}{2!}+\frac{(kcT(y-1))^3}{3!}\\
  -e^{4kcT(y-1)}=-1-4kcT(y-1)-\frac{(4kcT(y-1))^2}{2!}-\frac{(4kcT(y-1))^3}{3!}\\
  -3e^{2kcT(y-1)}=-3-6kcT(y-1)-3\frac{(2kcT(y-1))^2}{2!}-3\frac{(2kcT(y-1))^3}{3!}\\3e^{3kcT(y-1)}=3+9kcT(y-1)+3\frac{(3kcT(y-1))^2}{2!}+3\frac{(3kcT(y-1))^3}{3!}
  \end{eqnarray}
  Adding  the above four equations gives
  \begin{eqnarray*}
  	\left( e^{ kcT (y-1) }  -e^{  4kcT (y-1) }-3e^{  2kcT (y-1)}+3e^{ 3 kcT (y-1)} \right) =\\(1-1-3+3)+kcT(y-1)(1-4-6+9)+\frac{(kcT(y-1))^2}{2!}(1-16-12+27)\\+\frac{(kcT(y-1))^3}{3!}(1-64-24+81)\\ 
  	= -(kcT(y-1))^3\\
  	= (kcT(1-y))^3 
  \end{eqnarray*}
  Hence,  
  \begin{eqnarray*}
  	\langle C_{Pr} \rangle_n&&= \frac{4!C (kcT)^{n+1}}{6(kc)^n} \int_0^{1}   \left( e^{ kcT (y-1) }  -e^{  4kcT (y-1) }-3e^{  2kcT (y-1)}+3e^{ 3 kcT (y-1)} \right)y^n dy \nonumber\\ 
  	&&= \frac{4!C (kcT)^{n+1}}{6(kc)^n} \int_0^{1} (kcT(1-y))^3 * y^n dy \nonumber\\ 
  	&&= \frac{4!C (kcT)^{n+4}}{6(kc)^n} \int_0^{1} (1-y^3-3y+3y^2)  *   y^{n} dy, \quad kcT<<1\nonumber\\
  	&&= \frac{4!C (kcT)^{n+4}}{6(kc)^n}\left(\frac{1}{n+1}-\frac{1}{n+4}-\frac{3}{n+2}+\frac{3}{n+3}\right)\\
  	&&= \frac{4!C (kcT)^{n+4}}{6(kc)^n}\left(\frac{n+4-n-1}{(n+1)(n+4)}+\frac{-3n-9+3n+6}{(n+2)(n+3)}\right)\\
  	&&= \frac{4!C (kcT)^{n+4}}{6(kc)^n}\left(\frac{3}{(n+1)(n+4)}-\frac{3}{(n+2)(n+3)}\right)\\
  	&&= \frac{4!C (kcT)^{n+4}}{2(kc)^n}\left(\frac{1}{(n+1)(n+4)}-\frac{1}{(n+2)(n+3)}\right)\\
  	&&= \frac{4!C (kcT)^{n+4}}{2(kc)^n}\left(\frac{(n+2)(n+3)-(n+1)(n+4)}{(n+1)(n+4)(n+2)(n+3)}\right)\\
  	&&= \frac{4!C (kcT)^{n+4}}{2(kc)^n}\left(\frac{n^2+5n+6-n^2-5n-4}{(n+1)(n+4)(n+2)(n+3)}\right)\\
  	&&= \frac{4!C (kcT)^{n+4}}{(kc)^n(n+1)(n+2)(n+3)(n+4)}\\
  \end{eqnarray*}
  
Similarly,
  \begin{eqnarray*}
  	\langle C_{Pr}^2 \rangle_n
  	&&= \frac{4!C^2 (kcT)^{2n+4}}{(kc)^{2n}(2n+1)(2n+2)(2n+3)(2n+4)}\\
  \end{eqnarray*}
  
  Hence
  \begin{eqnarray*}
  	\frac{\sqrt{\langle C_{Pr}^2 \rangle_n -\langle C_{Pr} \rangle^2_n  }}{c\frac{d \langle C_{Pr} \rangle_n}{dc}}=\\
  	\frac{C*(kcT)^{n+2}}{kc^n} \sqrt{  \frac{4!}{(2n+1)(2n+2)(2n+3)(2n+4)}-\frac{4!*4!(kcT)^{4}}{(n+1)^2(n+2)^2(n+3)^2(n+4)^2}}*\\\frac{(kc)^n(n+1)(n+2)(n+3)(n+4)}{4*4!C (kcT)^{n+4}}\\
  	\\=
  	\sqrt{  \frac{4!}{(2n+1)(2n+2)(2n+3)(2n+4)}-\frac{4!*4!(kcT)^{4}}{(n+1)^2(n+2)^2(n+3)^2(n+4)^2}}*\frac{(n+1)(n+2)(n+3)(n+4)}{4*4!(kcT)^{2}}\\
  	\\=
  	\sqrt{  \frac{(n+1)^2(n+2)^2(n+3)^2(n+4)^2}{4!(2n+1)(2n+2)(2n+3)(2n+4)}-(kcT)^{4}}*\frac{1}{4(kcT)^{2}}\\
  	\\=
  	\frac{1}{4} \sqrt{  \frac{(n+1)^2(n+2)^2(n+3)^2(n+4)^2}{4!(kcT)^{4}(2n+1)(2n+2)(2n+3)(2n+4)}-1}
  	\\=
  	\frac{1}{4} \sqrt{  \frac{(n+1)(n+2)(n+3)^2(n+4)^2}{4!4(kcT)^{4}(2n+1)(2n+3)}-1}
  \end{eqnarray*}

 \section*{  “Data Availability Statement}
 There is no data associated with this manuscript.

\section*{Author contribution statement}
Both authors contributed equally to the manuscript
\end{document}